\documentclass[12pt]{iopart}
\usepackage{graphicx}
\usepackage{iopams}  
\begin{document}

\title[$A/\gamma^2$ relation in Yb-based systems]
{Deviation from the Kadowaki-Woods relation in Yb-based 
intermediate-valent systems}

\author{N. Tsujii$^1$\footnote  
{To
whom correspondence should be addressed (TSUJII.Naohito@nims.go.jp)},
 K. Yoshimura$^2$ and K. Kosuge$^2$}

\address{$^1$National Institute for Materials Science,
Sengen1-2-1, Tsukuba, Ibaraki 305-0047, Japan}

\address{$^2$Graduate School of Science, Kyoto University,
Kyoto 606-8502, Japan}

\begin{abstract}
The $T^2$ coefficient of electrical resistivity, $A$,
is compared with the electronic specific-heat coefficient,
$\gamma$, for a number of Yb-based compounds.
It is revealed that many systems including
YbCuAl, YbInCu$_4$, YbAl$_3$ and YbCu$_5$ show $A/\gamma^2$ values
close to 0.4$\times$10$^{-6}$ $\rm\mu\Omega{}cm (mol\cdot K/mJ)^2$,
which is remarkably smaller than that known as the Kadowaki-Woods relation,
$A/\gamma^2$ = 1.0$\times$10$^{-5}$ $\rm\mu\Omega{}cm (mol\cdot K/mJ)^2$.
Empirically, the compounds with the smaller $A/\gamma^2$ values
appear to show weak intersite magnetic-correlation
and/or to have almost fully-degenerated ($J$ = 5/2 or 7/2) ground states.

\end{abstract}


\submitto{\JPCM}

\maketitle

\section{Introduction}
The low-temperature properties of many U- and Ce-based intermetallic compounds
are well described by Fermi-liquid state~\cite{Stewart}.
Here, the specific heat $C$ and the electrical resistivity $\rho$
as a function of temperature $T$ vary as
$C \propto \gamma{}T$ and $\rho \propto AT^2$,
where $\gamma$ and $A$ are constant.
The coefficients $\gamma$ and $A$ are related to the
electron effective mass 
 $m^*$ as $\gamma \propto m^*$ and $A \propto (m^*)^2$, respectively.
The ratio $A/\gamma^2$ therefore does not depend on $m^*$.
In fact, Kadowaki and Woods
showed that many U- and Ce-based compounds
show a universal relation, $A/\gamma^2$ = 1.0$\times$10$^{-5}$
$\rm\mu\Omega$cm (mol$\cdot$K/mJ)$^2$~\cite{Kadowaki}.
Furthermore, this relation has proved to be applicable to $d$-electron
systems with large $\gamma$,
for example, A15-type intermetallics like V$_3$Si~\cite{Miyake}, 
nearly-ferromagnetic itinerant-electron systems like YCo$_2$~\cite{Gratz1}, 
and metallic oxides close to metal-insulator transition
like V$_2$O$_3$ under high pressure~\cite{Miyake} and LiV$_2$O$_4$~\cite{Urano}.
These observations demonstrate that the Fermi-liquid state
is a general feature for a wide range of correlated-electron systems.

Large $\gamma$ has also been reported in a number of Yb-based intermetallics
~\cite{Fisk}.
One can naturally expects that the Kadowaki-Woods relation,
$A/\gamma^2$ = 1.0$\times 10^{-5}$$\rm\mu\Omega$cm (mol$\cdot$K/mJ)$^2$, 
also holds for these Yb compounds.
In several Yb-based compounds, however, the $A/\gamma^2$ values
have been reported to be remarkably smaller than that of
the Kadowaki-Woods relation.
For YbCu$_4$Ag~\cite{Graf1,Graf2}
and YbCu$_{5-x}$Ag$_x$~\cite{Tsujii1,Tsujii2},
the $A/\gamma^2$ values are 
0.1-0.6$\times 10^{-6}$$\rm\mu\Omega$cm(mol$\cdot$K/mJ)$^2$,
nearly two orders of magnitude smaller than that for the 
Kadowaki-Woods relation.
It is also reported that the $A/\gamma^2$ 
for YbCu$_{4.5}$~\cite{Alami98,Alami98b}
and YbNi$_2$Ge$_2$~\cite{Knebel} is close to this value. 
In addition, this value well coincides with the ratio for the
transition metals such as Re, Os, Ni, and so on~\cite{Miyake,Rice}.

This deviation of $A/\gamma^2$ value from the Kadowaki-Woods relation
can pose quite an interesting subject to the Fermi-liquid theory,
if it is a general phenomena for a class of materials.
However, there have been no systematic studies on the $A/\gamma^2$
values for Yb-based compounds.
We have therefore investigated the $A/\gamma^2$ relation
in several Yb-based compounds.
We have measured the electrical resistivity of the Yb-based compounds,
YbAl$_2$, YbInAu$_2$, YbInCu$_4$, and YbCuAl.
These systems are typical examples of Yb-based
intermediate-valent or heavy-fermion compounds.
We also have examined the
$A/\gamma^2$ values for several other compounds
using the values reported in literature.

\section{Experimental}

Polycrystalline samples of YbAl$_2$ and YbCuAl
have been prepared by argon arc melting and subsequent annealing.
To compensate the loss of Yb due to evaporation, slightly
additional amount of Yb was used in synthesis. 
Polycrystalline sample of YbInAu$_2$ was prepared by melting the elements in
a BN crucible placed in an evacuated silica tube, and slow cooling.
For YbInCu$_4$, a single crystal sample was prepared by an In-Cu
flux technique~\cite{Sarrao}.
Powder x-ray diffraction
shows that single phased samples were obtained for all the compounds.
Electrical resistivity was measured by a standard dc four probe
technique in the temperature range between 4.2 and 300 K.

\section{Results}

Figure 1 shows the electrical resistivity $\rho$  
as a function of temperature $T$.
The residual resistivity $\rho_0$
of these samples are almost the same as or smaller than those
reported in the literature
~\cite{Alami98,Sarrao,Mignot81,Cattaneo86,Daal74,Havinga73}.
$\rho$ of YbInCu$_4$ decreases abruptly at 40 K due to the valence transition
of Yb~\cite{Sarrao}. The sharp transition indicates that our crystal 
is well ordered~\cite{Lawrence1}. 
YbCuAl shows relatively rapid decrease below approximately 20 K.
This temperature roughly corresponds to the characteristic temperature
$T^*$, below which the coherence between Kondo singlets develops and 
the system enters the Fermi liquid regime~\cite{Lavanga}. 
For YbInAu$_2$, $T^*$ is of the order of 70 K.
For YbAl$_2$, $\rho$ does not show a distinct temperature
dependence up to 300 K. This indicates that $T^*$
of YbAl$_2$ is above the room temperature.

In Fig. 2, $\rho-\rho_0$ are plotted as a function of $T^2$.
For YbCuAl, YbInAu$_2$, and YbAl$_2$, $\rho-\rho_0$ is described
by the $T^2$ power law at low temperatures,
indicating the evolution of Fermi-liquid state.
For YbInCu$_4$, $\rho-\rho_0$ deviates from the $T^2$ dependence
above 20 K, and the data up to 30 K are well fitted with the 
correction of $T^5$ dependence, as is shown in Fig. 2 (b).
This suggests that the conventional electron-phonon scattering
is important in YbInCu$_4$.

The values of the $T^2$ coefficient $A$ have been obtained
by fitting the data at low temperatures, and are listed in table 1.
In these $A$ values, errors of the order of 20\% can exist
due to the inaccuracy in estimating the sample dimension.
For YbAl$_2$ and YbInCu$_4$, however, the error in $A$ can be much larger because of 
the small value of $A$ and the presence of the $T^5$ term.

The $A$ values are plotted against $\gamma$
on the Kadowaki-Woods plot
~\cite{Kadowaki,Miyake,Gratz1}, as is shown in Fig. 3.
We also plot $A$ and $\gamma$ values of several systems reported
in literatures. The values of $A$ and $\gamma$ are listed in
table 1.

In Fig. 3, one can see that the Kadowaki-Woods relation,
$A/\gamma^2$ = 1.0$\times$10$^{-5}$ $\rm\mu\Omega$cm(mol$\cdot$K/mJ)$^2$, is almost
preserved for YbNi$_2$B$_2$C, YbRh$_2$Si$_2$, Yb$_2$Co$_3$Ga$_9$, CeNi, and YbInAu$_2$.
On the other hand, it is apparent that several systems 
show a significant deviation from the relation. 
In particular, there appears to exist another 'universal' relation; 
$A/\gamma^2 \simeq 0.4\times{}10^{-6} \rm\mu\Omega$cm (mol$\cdot$K/mJ)$^{2}$,
for YbCu$_{5-x}$Ag$_x$, YbCu$_{4.5}$, YbCuAl, YbNi$_2$Ge$_2$,
YbAl$_3$, YbInCu$_4$, YbCu$_4$Al, CeNi$_9$Si$_4$ and CeSn$_3$.
Extrapolation of this line well coincides 
with the $A/\gamma^2$ value for the transition metals such as
Pd or Re~\cite{Rice}.
As for YbAl$_2$  and YbCu$_2$Si$_2$, the $A/\gamma^2$ ratio appears to
be situated between the two lines.

\section{Discussion}

We have shown in the preceding section that
many Yb-based and several Ce-based compounds show 
nearly two times smaller values of $A/\gamma^2$
than that for the Kadowaki-Woods relation.
In this section, we discuss on the possible origin of the deviation
from the Kadowaki-Woods relation.

At first, it must be remarked that the $A$ value depends not only on $T_{\rm K}$
but also on many factors, including
anisotropy, carrier concentration, Fermi-surface topology, band structures,
and site disorder.
In particular, the effect of site disorder may be important
as is pointed by several authors, 
since the effect can suppress $A$ through the Kondo-hole effect
~\cite{Alami98,Alami98b,Knebel}.
Here, the site disorder gives the additional resistivity:
$\rho^{\rm disorder}$ = $\rho_0-BT^2$~\cite{KH},
which results in the reduced $T^2$ coefficient of $A-B$.
However, although this mechanism may be important in some systems, 
it is unlikely that the Kondo-hole effect
can explain all the small $A/\gamma^2$ values,
especially almost universal value, 
$A/\gamma^2$ = 0.4$\times$10$^{-6}$ $\rm\mu\Omega{}cm (mol\cdot K/mJ)^2$.
Kondo-hole effect should result in a random distribution of the $A/\gamma^2$ values.
Moreover, the additional term,
$\rho^{\rm disorder}$ = $\rho_0-BT^2$, 
has been observed in only limited systems,
such as YbInAu$_2$ under high pressures~\cite{Alami98}.
Thus, we believe that the Kondo-hole effect is not the crucial origin
for most of the small $A/\gamma^2$ values.
As for the other effect such as carrier concentration or band structures,
although these effect must be taken into consideration,
it is unlikely that these effect can explain the smaller 'universal' relation.

We next consider this phenomenon relating with the three mechanisms;
(i) single-body band effect, 
(ii) intersite magnetic correlation,
and (iii) the ground state degeneracy.

\subsection{Single-body band effect}

Miyake {\it et al.} have considered theoretically 
the effect of many-body correlations in 
heavy-fermion systems, where conduction electrons have a
strong frequency-dependence in its self energy.
They have suggested that the 
many-body effect can enhance the $A/\gamma^2$ value 
by about 25 times larger than the case of single-body band,
and this would explain the 25 times different $A/\gamma^2$
between the transition metals and the heavy-fermion systems~\cite{Miyake}.

Similar argument may be applicable for YbAl$_2$,
of which $\gamma$ is relatively small.
In this compound, the many-body and the single-body effects
can be comparable, which may
reduce the $A/\gamma^2$ to the same order as that in transition metals.
On the other hand, for YbCuAl, YbCu$_4$Ag, YbCu$_5$ or CeNi$_9$Si$_4$,
these systems exhibit much larger $\gamma$ (100-600 mJ/molK$^2$)
than those of transition metals.
This indicates that DOS of the conduction band
is enhanced through the many-body effect, i.e., the Kondo-lattice formation.

Even in YbAl$_3$, for which the $\gamma$ value (45 mJ/mol$\cdot$K$^2$)
may not be so large, the de-Haas van Alphen effect measurements
have revealed that the Fermi surface of YbAl$_3$ is strongly renormalized
due to many-body effect of the localized 4$f$ moment and the conduction
electrons~\cite{Ebihara2}.
We therefore suppose that the single-body band effect is not dominating
for these systems.

\subsection{Intersite magnetic correlation}
As is pointed in ref.~\cite{Kadowaki}, many
U-based compounds on the Kadowaki-Woods line 
exhibit strong intersite magnetic-correlation,
including magnetic ordering in UPt, UGa$_3$ and UIn$_3$,
and distinct spin-fluctuations in UAl$_2$ and UPt$_3$.
Magnetic correlation due to the Ruderman-Kittel-Kasuya-Yosida (RKKY) 
interaction is also considered to be important
for those Ce- and Yb-based compounds such as
CeAl$_3$~\cite{CeAl3,CeAl3b}, CeCu$_2$Si$_2$~\cite{Ishida99},
CeB$_6$~\cite{Sato}, CeCu$_6$, CeRu$_2$Si$_2$~\cite{Mignod},
YbRh$_2$Si$_2$~\cite{Trovarelli,Gegenwart}, and so on.
For $d$-electron systems such as LiV$_2$O$_4$~\cite{Lee}, YCo$_2$~\cite{Yoshimura}
and (Y,Sc)Mn$_2$~\cite{Shiga2},
the presence of strong magnetic correlation has been demonstrated
by neutron scattering or NMR.
It is notable that these compounds include both nearly-ferromagnetic
(UAl$_2$~\cite{Takagi}, YCo$_2$~\cite{Yoshimura} etc.)
and nearly-antiferromagnetic systems 
(CeCu$_6$~\cite{Mignod}, CeCu$_2$Si$_2$~\cite{Ishida99} etc.).

On the other hand, for the systems with the smaller $A/\gamma^2$ value,
magnetic correlation does not appear to be important.
Physical properties are well explained using a single-impurity model
in YbCu$_5$~\cite{Tsujii2,Tsujii3}, YbCu$_{5-x}$Ag$_x$
~\cite{Tsujii1,Michor02,Hauser}, 
YbCu$_4$Ag~\cite{Rossel,Besnus}, YbCuAl~\cite{Hewson}, CeNi$_9$Si$_4$~\cite{MichorLT} 
and CeSn$_3$~\cite{Schlottmann}.
Single-impurity character is also demonstrated by neutron scattering experiments
for YbCuAl~\cite{Murani85} and YbInCu$_4$~\cite{Lawrence97}.
Therefore in these compounds, intersite magnetic correlation can be neglected.
For YbCu$_{4.5}$, magnetic ordering is not observed even at 235 kbar
down to 50 mK~\cite{Link95}, suggesting the weak RKKY interaction.

This remarkable contrast in the strength of magnetic interactions
implies that there exist qualitatively different scattering mechanisms
in the electrical resistivity.
For the case with strong magnetic interactions, $A$ and $\gamma$ are
enhanced both by the large density of state (DOS) at the Fermi energy ($E_{\rm F}$)
and by the strong spin-fluctuations.
In contrast, for the single impurity-like case, the large $A$
is solely attributed to the DOS at $E_{\rm F}$.
One may hence consider that this difference is the possible origin 
of the two 'universal' $A/\gamma^2$ relations. 

However, this tendency is inconsistent with the theoretical calculations
based on the spin fluctuation theory~\cite{Takimoto,Continentino}.
Takimoto {\it et al.} have shown that the value of $A/\gamma^2$ is
almost constant from the pure Kondo regime (single-impurity limit)
to almost magnetically instability limit.
The value of $A/\gamma^2$ they estimated is close to the Kadowaki-Woods
relation~\cite{Takimoto}.
Similar result is also reported by Continentino,
who have shown that the $A/\gamma^2$ value is independent
of the distance from magnetic instability~\cite{Continentino}.

It should also be noted that there exist some exceptions of the above tendency.
In CePd$_3$, intersite magnetic correlation is considered 
to be of minor importance~\cite{Murani96}.
For YbInAu$_2$ or Nb$_3$Sn, no evidence of strong magnetic-correlation is
reported. However, $A/\gamma^2$ values in these systems are close to
the Kadowaki-Woods relation.
On the other hand, 
Pd and Pt show the small $A/\gamma^2$ value, 
though they are well-known examples of nearly-ferromagnetic metals~\cite{Takigawa}.
Therefore, these findings may not justify the above scenario, in which 
strong intersite magnetic-correlation enhances $A/\gamma^2$ values.

\subsection{Ground state degeneracy}
One may notice that most of the systems with the smaller $A/\gamma^2$ values
in Fig. 3 are classified as the intermediate-valent compounds,
where crystal-field splitting, $\Delta_{\rm CF}$, is of minor importance, and 
the full degeneracy $N$ for Yb$^{3+}$ or Ce$^{3+}$ ion is 
almost preserved~\cite{Brandt}.
In fact, physical properties of YbCu$_4$Ag, YbCuAl, CeSn$_3$ and CeNi$_9$Si$_4$
are explained by the impurity model for $N$ = 8 (Yb$^{3+}$) or $N$ = 6 (Ce$^{3+}$),
as is noted in the preceding subsection.
For YbAl$_3$ and YbAl$_2$, Shimizu {\it et al.} have suggested by the $^{171}$Yb-NMR
that the $N$ = 8 ground state is well conserved in these compounds~\cite{Shimizu}.
Even in YbCu$_5$, which exhibits a large $\gamma = 550$ mJ/mol$\cdot$K$^2$,
$\Delta_{\rm CF}$ is considered to be comparable with the Kondo temperature $T_{\rm K}$,
and the low temperature properties are characterized by $N$ = 4~\cite{Michor02,Hauser}

On the contrary, the systems such as CeCu$_6$, CeAl$_3$ and CeCu$_2$Si$_2$ 
are considered to be affected by the crystal-field effect,
which reduce the ground state degeneracy down to $N$ = 2~\cite{Brandt}.

The difference in $N$ is related to the magnitude
of the magnetic moment which participate in the Fermi-liquid formation.
It can therefore affect on the $A/\gamma^2$ values.
We note that the difference between $N \geq$ 4 and $N \leq$ 3 states 
can be crucial.
Theoretical calculation predicts that the DOS has a peak situated
above $E_{\rm F}$ for the case of $N \geq$ 4, 
whereas the peak is positioned just at $E_{\rm F}$ for $N \leq$ 3~\cite{Rajan}.
Similarly, magnetic excitation spectra also 
show inelastic structure for large $N$~\cite{Kuramoto,Cox85}.
Hence, we suppose that in those Yb- and Ce-based intermediate-valent systems,
their largely-degenerated states ($N \geq 4$) develop the excitation spectra
with qualitatively different shape from that in those
heavy-fermion systems with small $N$ ($\leq$ 3),
resulting in the significant difference in the $A/\gamma^2$ values.
There have been no theories for the $N$ dependence of
$A/\gamma^2$ value as far as we know.
Theoretical investigation for the relation between $N$ and $A/\gamma^2$
is desired~\cite{LawrencePC}.

However, some exceptions for this understanding should be noted.
For Yb$_2$Co$_3$Ga$_9$~\cite{Dhar01},
the full degeneracy for Yb$^{3+}$ is considered
to be preserved, though they exhibit the same order of $A/\gamma^2$
as that for the Kadowaki-Woods relation.
In addition, it is unclear whether all the U-based compounds 
on the Kadowaki-Woods line can be understood in terms of
$N \leq 3$ ground states.
In these U-compounds, the enhanced low-energy excitations due to 
intersite magnetic correlation may also be responsible.

We also note that CePd$_3$ is again exceptional.
This compound is a typical intermediate-valent 
compound. The full degeneracy $N$ = 6
is considered to be preserved, and the intersite magnetic correlation
is also of minor importance~\cite{Murani96}.
Nevertheless, this compound show the same order of $A/\gamma^2$ value
as that of heavy-fermion compounds.
However, it should be remarked that 
the carrier concentration of CePd$_3$ is unusually small 
($\sim$ 0.3 per unit cell)~\cite{Webb}, and hence this compound
would be close to Kondo-insulators.
This situation cannot allow us to make a naive comparison
for CePd$_3$.
\\
\\
Finally, an important suggestion is given experimentally from 
the electrical resistivity measurement under pressures.
Knebel {\it et al.}~\cite{Knebel} 
have examined the $AT_{\rm max}^2$ of YbNi$_2$Ge$_2$ under pressure,
where $T_{\rm max}$ is the maximum temperature of the resistivity
adhering to a $T^2$ power-law,
and should be proportional to $\gamma^{-1}$.
The value of $AT_{\rm max}^2$ is therefore a measure of $A/\gamma^2$.
They have found that $AT_{\rm max}^2$ increases by a factor about 25 times
under high pressures.
Similar variation of $AT_{\rm max}^2$ under pressure has also been reported
for CeCu$_2$Si$_2$~\cite{Jaccard99}.
These phenomena are most likely related to either of the change
of the intersite magnetic correlation or that of the ground state degeneracy.
Hence, performing similar experiments on much more systems
may eventually reveal the underlying origin of the two 'universal' $A/\gamma^2$
relations.

In addition, the effect of disorder must be carefully taken into consideration.
If the disorder makes little effect on the host-material properties,
its effect on the resistivity would be described by the Kondo-hole effect, 
which reduces $A$.
On the contrary, in most of systems, disorder is considered to affect the whole range of electrical
properties in the Kondo-lattice formation.
It would reduce the characteristic temperature $T^{*}$, 
which results in the enhanced $A$ values.
This may be the case in YbInAu$_2$, where site disorder
easily occur because of its crystal structure.
Therefore, the $A/\gamma^2$ values in several systems may not be intrinsic.
Experimental works using high quality samples are particularly important.

\section{Summary}
We have measured the electrical resistivity
and have examined the $A/\gamma^2$ values of YbAl$_2$, YbInAu$_2$, 
YbInCu$_4$ and YbCuAl.
We have also plotted $A$ vs. $\gamma$ for several systems reported 
in literature.
The result reveals that the $A/\gamma^2$ values are not unique among materials,
but varies in the range of 0.4$\times$10$^{-6}$ - 1.0$\times$10$^{-5}$
$\rm\mu\Omega$cm(mol$\cdot$K/mJ)$^2$.
In particular, it has been found that there exists another 'universal' relation;
$A/\gamma^2$ = 0.4$\times$10$^{-6}$ $\rm\mu\Omega$cm(mol$\cdot$K/mJ)$^2$,
for a number of systems including YbCuAl, YbAl$_3$, YbInCu$_4$, YbCu$_4$Al as well as 
YbCu$_5$, YbCu$_{4.5}$, YbCu$_4$Ag, YbNi$_2$Ge$_2$, CeNi$_9$Si$_4$ and CeSn$_3$.
This value of $A/\gamma^2$ is about 25 times smaller than that known in 
the Kadowaki-Woods relation;
$A/\gamma^2$ = 1.0$\times$10$^{-5}$ $\rm\mu\Omega$cm(mol$\cdot$K/mJ)$^2$.

To explain this deviation,
several possible mechanisms have been discussed.
The Kondo-hole effect or the single-body-band effect 
may be applicable for some systems.
However, these two mechanisms cannot explain the almost 'universal' nature
of the smaller $A/\gamma^2$ values.
Thus, we have considered the tendency of these compounds empirically,
and have focused on the intersite magnetic correlation
and the ground state degeneracy.

As for the former,
we have pointed that most of the systems with the smaller $A/\gamma^2$ value
exhibit single-impurity like behavior. 
This suggests that the intersite magnetic interactions are negligible
in these systems.
In contrast, many systems on the Kadowaki-Woods
line are known to show strong magnetic correlation.
This tendency is valid for many compounds plotted
in Fig. 3. However, this interpretation is inconsistent
with the theoretical results based on the spin-fluctuation theory.

As for the latter, it has been found that the ground state degeneracy $N$
differs largely between the systems with the smaller $A/\gamma^2$
and those on the Kadowaki-Woods line.
Almost all the systems with the smaller $A/\gamma^2$ values
(except for $d$-electron systems) show
largely degenerated ground state; i.e., $N \geq$ 4.
On the other hand, those on the Kadowaki-Woods line are considered to have
$N \leq 3$ state due to the crystal-field splitting.
Since this difference can affect on the magnitude of the magnetic moment
as well as the shape of the excitation spectra,
the $A/\gamma^2$ values may also vary depending on $N$.

For both the cases, i.e., the intersite magnetic correlation and 
the ground state degeneracy, 
our interpretations are only empirical.
Hence, theoretical confirmations are strongly desired.

\section*{Acknowledgments}
Authors gratefully acknowledge T Takimoto, S Kambe,
K Miyake, J M Lawrence, G Hilscher, H Kitazawa and H Suzuki
for valuable discussion and suggestions.

\pagebreak
\begin{table}
 \begin{center}
  \begin{tabular}{llllll}   \hline
Compounds                            &  $A$ & $\gamma$ & $A/\gamma^2$  & Magnetic    & Ground state \\ 
                                     &      &          &               & correlation & degeneracy   \\ \hline
YbCu$_5$ ~\cite{Tsujii2}               & 0.15    & 550   & 0.05 & w & 4~\cite{Michor02} \\
YbCu$_{4.5}$Ag$_{0.5}$ ~\cite{Tsujii1} & 0.064   & 380   & 0.04 & w & 6~\cite{Hauser} \\
YbAgCu$_4$ ~\cite{Tsujii1}             & 0.023   & 210   & 0.05 & w & 8~\cite{Rossel,Besnus}  \\
YbCu$_{4.5}$ ~\cite{Fisk,Link95}       & 0.1     &  600  & 0.03 & w &    \\
YbCuAl ~\cite{Fisk}                    & 0.068 * & 260   & 0.1  & w ~\cite{Hewson,Murani85} & 8 \\
CeNi$_9$Si$_4$ ~\cite{MichorLT}        & 0.017   & 156    &  0.07 & w &  6 \\
YbNi$_2$Ge$_2$ ~\cite{Knebel,Budko99}  & 0.0078  & 136    &  0.04 &   &  \\
YbCu$_4$Al ~\cite{Bauer97,Bauer00}     & 0.0025  & 50-100 & 0.03-0.1  &  &  \\
YbAl$_3$ ~\cite{Fisk,Ebihara1}   & 5$\times$10$^{-4}$ & 45   &  0.06 & w ~\cite{Shimizu} & 8 ~\cite{Shimizu} \\
CeSn$_3$ ~\cite{Kadowaki}        &        &      & $\sim$ 0.03   & w & 6 ~\cite{Schlottmann} \\
YbInCu$_4$ ~\cite{Pillmayr}      & 6$\times$10$^{-4}$ * & 50 & 0.03 & w ~\cite{Lawrence97} & 8 ? \\
YbAl$_2$ ~\cite{Fisk}            & 4$\times$10$^{-4}$ * & 17 & 0.14 & w ~\cite{Shimizu} & 8 ~\cite{Shimizu} \\
Pd, Pt ~\cite{Rice}                 &          &    & 0.04 & s ~\cite{Takigawa} &  \\ \hline
Yb$_2$Co$_3$Ga$_9$ ~\cite{Dhar99}      & 0.054 & 112  & 0.43  & w ~\cite{Dhar01} & 8 ~\cite{Dhar01}\\
YbCu$_2$Si$_2$ ~\cite{Thompson87,Fisk}  & $\sim$0.04  & 135  &  $\sim$0.2 &     &     \\
CeNi ~\cite{Oomi,Sereni} & 0.013 & 60 & 0.36 & m ? ~\cite{Clementyev} &   \\
YbInAu$_2$ ~\cite{YbInAu2}           & 0.0070 * & 40   &  0.44    &     &     \\ \hline
CeCu$_6$ ~\cite{Kadowaki}     &       &   & $\sim$ 1.0  &  s ~\cite{Mignod} &  2 ?\\
UPt$_3$ ~\cite{Kadowaki}      &       &   & $\sim$ 1.0  &  s ~\cite{Koike} &   \\
LiV$_2$O$_4$ ~\cite{Urano}    & 2.0  & 200 & 5 &  s ~\cite{Lee} &  2  \\
YbRh$_2$Si$_2$ (0 T) ~\cite{Trovarelli,Gegenwart}   & 22 & 1700 & 0.8 & s  & 2 ?\\
YbRh$_2$Si$_2$ (6 T) ~\cite{Trovarelli,Gegenwart}& 1.0 & 300 & 1.1  & s  ~\cite{Ishida02} & 2 ?\\
YbNi$_2$B$_2$C ~\cite{Yatskar} & 1.2 & 530 & 0.43 & m ? & 2-3 \\
CeRu$_2$Si$_2$ ~\cite{Kambe96} & 0.62 & 350 & 0.51 & s  ~\cite{Mignod} & 2 ?\\
Sr$_3$Ru$_2$O$_7$ ~\cite{Ikeda} & 0.075 & 110 & 0.6 & s & 3 \\
CePd$_3$ ~\cite{Kadowaki} &     &   & 2.5 & w ~\cite{Murani96} & 6 ~\cite{Murani96}\\
V$_2$O$_3$ ~\cite{Miyake} &    &    &  4  & s &  3 \\
YCo$_2$ ~\cite{Gratz1} &        &   &     &  s ~\cite{Yoshimura} &     \\
(Y,Sc)Mn$_2$ ~\cite{Shiga1} & 0.2 & 80 & 3.1 & s ~\cite{Shiga2} &  \\ \hline
          &(* this work)
\end{tabular}
\end{center}
\vspace{1cm}
\caption{
List of the $T^2$ coefficient of electrical resistivity $A$ 
and the $T$-linear specific heat coefficient $\gamma$.
The units of $A$, $\gamma$, and $A/\gamma^2$ are
$\rm\mu\Omega$cm/K$^2$, mJ/mol$\cdot$K$^2$, and
10$^{-5}\rm\mu\Omega$cm(mol$\cdot$K/mJ)$^2$, respectively.
Here 'mol' means magnetic-ion mol.
The values of $A$ given in this work are marked by asterisk.
The character w, m, and s indicates that the magnetic correlation
is weak, mediate, and strong, respectively.
}
\end{table}%

\pagebreak
\begin{figure}[!tbp]
\begin{center}
 \includegraphics[width=10cm]{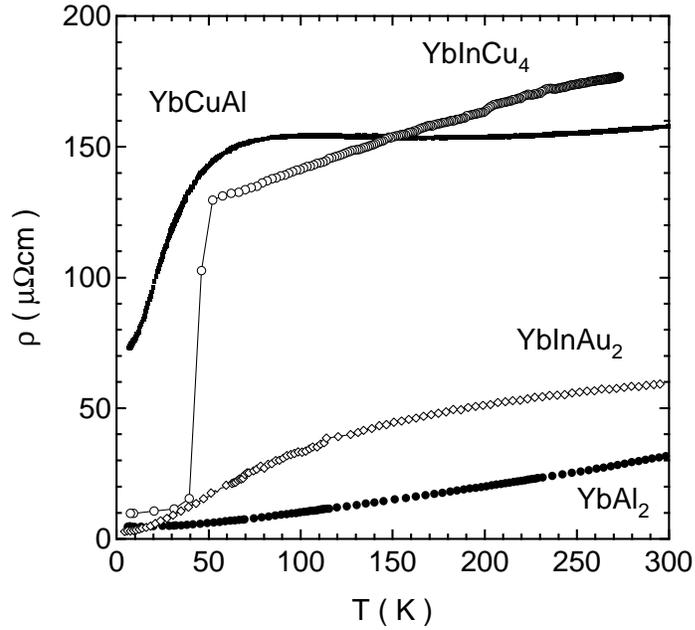}
\end{center}
\caption{
Electrical resistivity $\rho$ of YbAl$_2$, YbInAu$_2$, YbCuAl polycrystals
and flux-grown YbInCu$_4$ single crystal
as a function of temperature $T$.}
\end{figure}%

\begin{figure}[!tbp]
\begin{center}
 \includegraphics[width=14cm]{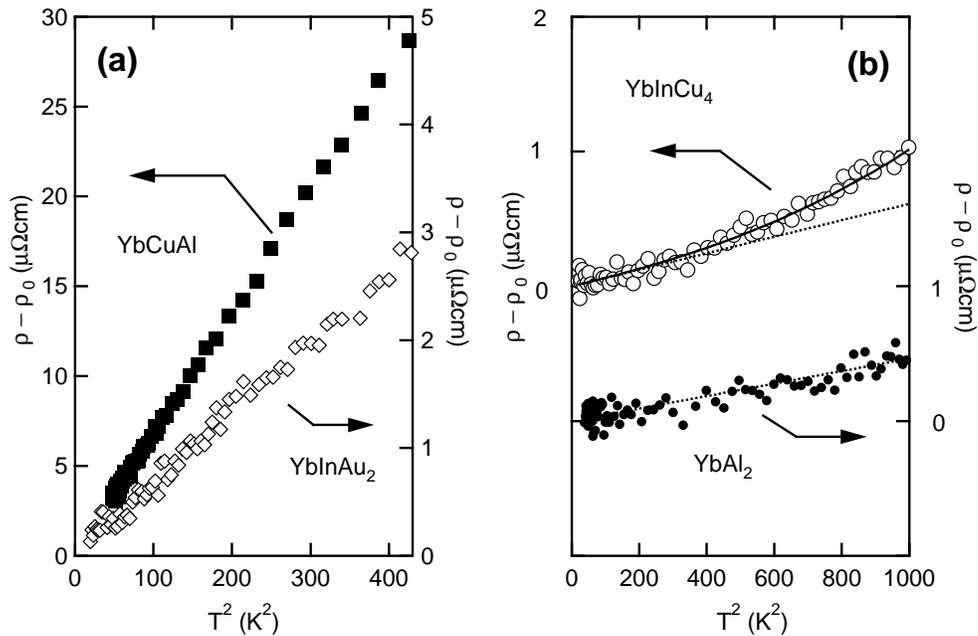}
\end{center}
\caption{
Electrical resistivity with the residual resistivity subtracted, $\rho - \rho_0$,
as a function of the square of temperature, $T^2$.
The dotted lines in (b) indicate the $T^2$ power law.
The solid line for YbInCu$_4$ is the result of fitting by the sum of
$T^2$ and $T^5$ terms.}
\end{figure}%

\begin{figure}[!tbp]
\begin{center}
 \includegraphics[width=12cm]{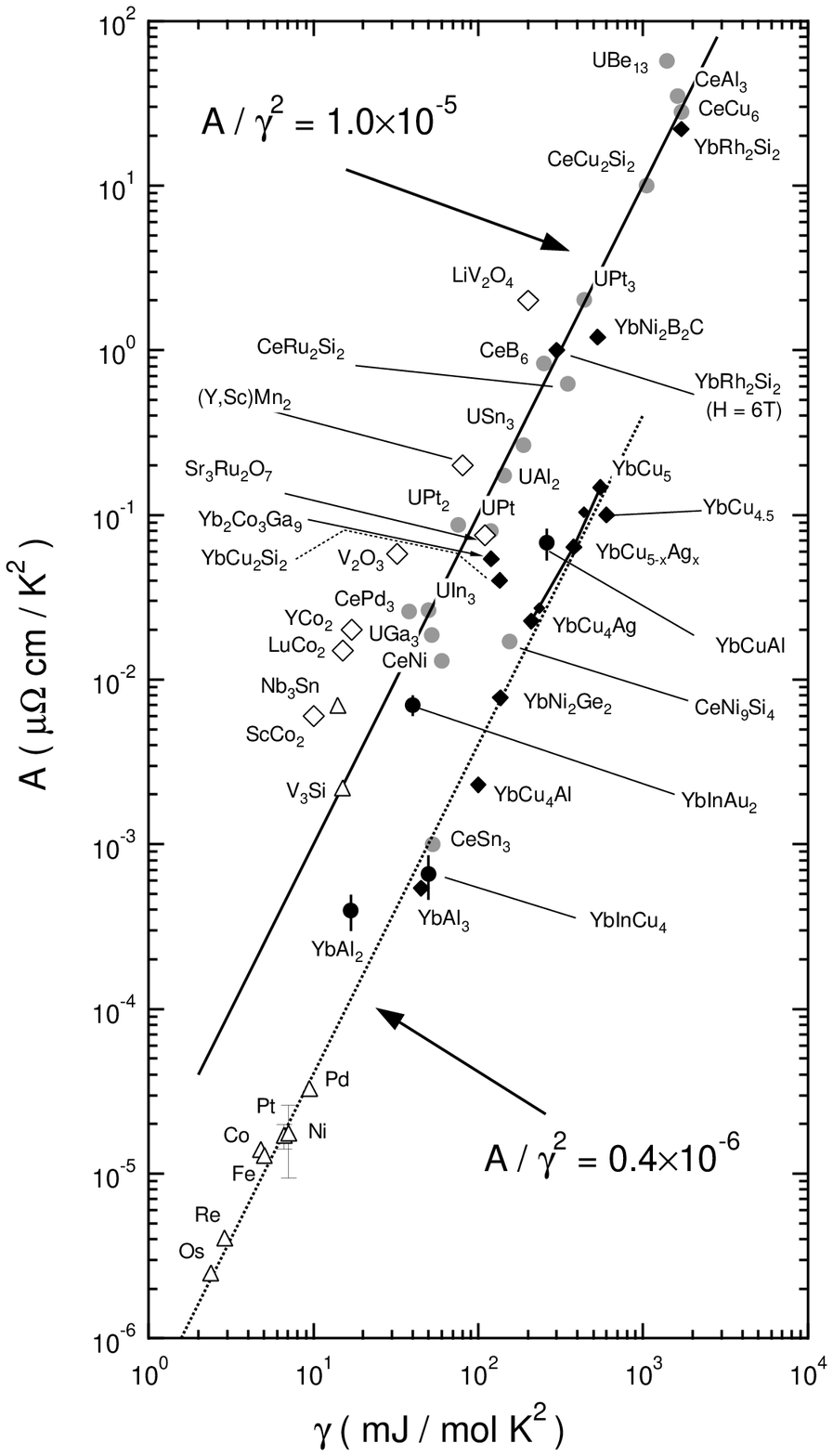}
\end{center}
\caption{
Plot of the $T^2$ coefficient of electrical resistivity $A$ vs. 
the $T$-linear specific heat coefficient $\gamma$.
Solid and dotted lines represent $A/\gamma^2$ = 1.0 $\times$10$^{-5}$
and 0.4 $\times$10$^{-6}$ $\rm\mu\Omega$cm(mol$\cdot$K/mJ)$^2$, respectively.
Data for Yb compounds are represented by black-filled symbols.
Black-filled circles ({\large $\bullet$}) indicate the data of
the present work.
Gray-filled symbols represent Ce- and U-based compounds.
$d$-electron based systems are indicated by open symbols.
For references, see table 1.}
\end{figure}%


\begin{thebibliography}{00}
\bibitem{LawrencePC}Lawrence J M has tentatively suggested that 
                    $A/\gamma^2$ can be a decreasing function of $N$
                    (private communication),
                   if we assume that $A$ = $\rho^{\rm unit}$/$T_{\rm K}^2$,
                   where $\rho^{\rm unit}$ is the unitarity limit described
                   as $\rho^{\rm unit} \propto N{\rm sin}^{2}(\pi/N)$, and that
                   $\gamma = (N-1)\pi{}k_{\rm B}/6T_{\rm K}$~\cite{Rajan}.
\end{thebibliography}
\end{document}